\documentclass{aa}  

\usepackage{graphicx}
\usepackage{txfonts}
\usepackage[usenames, dvipsnames]{color}

\begin{document}

   \title{Ionization memory of plasma emitters in a solar prominence}

   \titlerunning{Ionization memory of plasma emitters}

   \author{E. Wiehr\inst{1}\fnmsep\thanks{Deceased on 03 February 2025},
          H. Balthasar\inst{2},
          G. Stellmacher\inst{3},
          \and
          M. Bianda\inst{4}
          }

   \authorrunning{Wiehr et al.}
   
   \institute{Institut f\"ur Astrophysik, Universit\"at G\"ottingen,
              Friedrich-Hund-Platz 1, 37077 G\"ottingen, Germany\\
         \and
             Leibniz-Institut f\"ur Astrophysik Potsdam (AIP), An der Sternwarte 16,
                14482 Potsdam, Germany\\
             \email{hbalthasar@aip.de}
         \and
             Institute d'Astrophysique, 98 bis Boulevard Arago, 75014 Paris, France
         \and
             Istituto ricerque solari Aldo e Cele Dacc\'o (IRSOL), Universit\'a della
             Svizzera italiana, Via Patocchi 57, 6605 Locarno-Monti, Switzerland
             }

   \date{Received July 16, 2024; accepted March 07, 2025}

 
  \abstract
   { }
   {In the low-collisional, partially ionized plasma (PIP) of solar prominences, uncharged 
   emitters might show different signatures of magnetic line broadening than 
   charged emitters. We investigate if the
   widths of weak metall emissions in prominences exceed the thermal line broadening
   by a different amount for charged and for uncharged emitters.
   }
   {We simultaneously observe five optically thin, weak metall lines in the brightness 
    center of a quiescent prominence and compare their observed widths with the thermal 
     broadening.  
    }    
   {The inferred non-thermal broadening of the metall lines does not indicate systematic 
    differences between the uncharged Mg\,b$_2$ and Na\,D$_1$ and the charged 
    \ion{Fe}{ii} emitters, only \ion{Sr}{ii} is broader. 
    }
   {
   The additional line broadening of charged emitters is reasonably attributed 
   to magnetic forces. That of uncharged emitters can then come from their temporary 
   state as ion before recombination. Magnetically induced velocities will retain some 
   time after recombination. Modelling partially ionized plasmas then requires consideration
   of a memory of previous ionization states.
   }


   \maketitle
%

\section{Introduction}\label{sec:intro}

The study of partially ionized plasmas (PIP) has become increasingly important 
in recent years  
\citep{Khomenko2017, Ballester_etal2018, SolerBallester2022, Parenti_etal2024,
    Heinzel2024}.
A reduced collisional rate of the PIP enables a decoupling of charged and 
uncharged species which leads to different dynamical behavior of them. 
Higher flow velocities (line shifts) of ions are observed by
\citet{Khomenko_etal2016, WiehrEtal2019, WiehrEtal2021, ZapiorEtal2022}
in solar prominences, which are ideal objects to 
observe such effects with high spatial and temporal resolution. Drifts of ions 
relative to neutrals pose a particular problem for prominence support against 
gravity by magnetic forces since neutrals will sink through the magnetic structure 
if they are not in collisional equilibrium with charged particles.
\citet{Gilbert2002, Gilbert2007} report a depletion of helium in the high parts 
of filaments.
For prominences, we expect a certain coupling 
between the two, but weak enough to allow drifts between ions and 
neutrals. 

The decoupling of charged and uncharged species in a PIP
can also manifest itself in a different line broadening. 
Already \citet{Landman1981}
expects lines from ions to be broader than such from neutrals.
\citet{Ramelli_etal2012}
find for a quiescent 
prominence a \ion{He}{ii} line 1.5 times broader than a \ion{He}{i} 
line from the triplet system which, in turn, is 1.1 times broader than a 
\ion{He}{i} line from the singlet system. 
\citet{StellmacherWiehr2015}
find that the Mg\,b$_2$ line is 1.3 times broader than the ('forbidden')
inter-combination line from the magnesium triplet to the singlet system. 
\citet{StellmacherWiehr2017}
find a width excess of \ion{Sr}{ii} and 
\ion{Fe}{ii} lines relative to Na\,D and, respectively, He-singlet,
which they interpret by influences of magnetic forces. 
\citet{Gonzalez_etal2024}
observe the \ion{Ca}{ii}\,8542\,\AA{} line broader than
expected from He\,D$_3$ and H$\alpha$. Since strong ('chromospheric')
lines such as \ion{Ca}{ii}\,8542\,\AA{} and H$\alpha$ are optically thick 
in most parts of a prominence, the result is 
limited to the boundary between prominence and the corona where 'instabilities may take 
place... helping to increase the differential ion-neutral behavior...' 
\citep{Gonzalez_etal2024}.
\citet{Pontin_etal2020} investigate for coronal loops such line broadening in terms of magnetic
'braiding induced turbulence', but it is not fully clear if such an effect also applies in 
prominences.

\citet{Parenti_etal2024}
resume that 'we need the measurement of several spectral profiles 
from neutrals and low ionization state species under optical thin condition'.
This is the aim of the present study.
We observe weak metall lines which are optically thin 
throughout the whole prominence allowing to analyze the line broadening in 
the central prominence body. Such weak metall lines are rarely observed
in prominences; they are mostly drowned in the parasitic light of the aureole
and therefore require careful correction compared to bright 'chromospheric' lines.

\section{Observations and data reduction}\label{sec:obs}

\begin{figure}[ht!]
\includegraphics[width=0.465\textwidth]{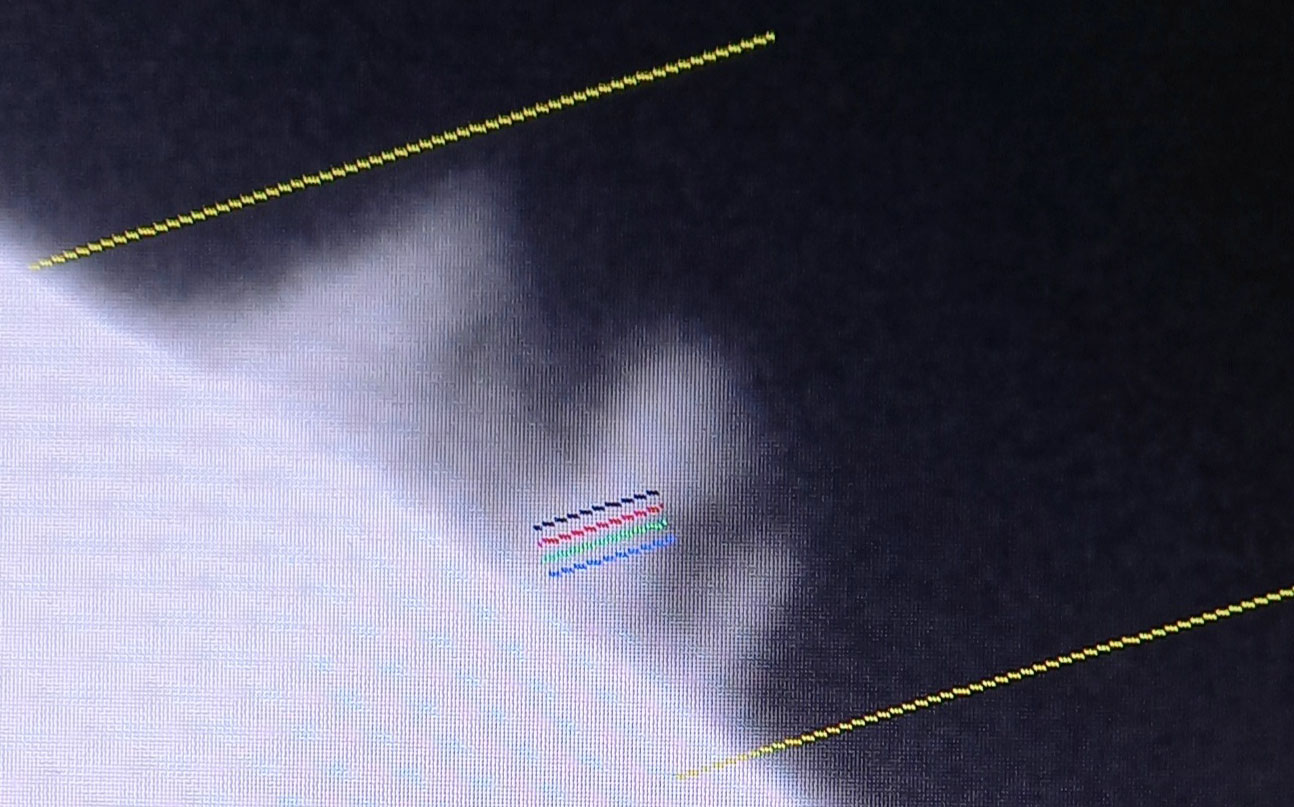}
\caption{H${\alpha}$ image of the prominence at the west limb, $52^\circ$ north,
  from June 6, 2022, (GONG archive); north direction is upwards; the long lines
  mark the direction of refraction at two slit positions aside the prominence used
  for the determination of parasitic light; the short lines mark slit positions
  reconstructed from the observed H$\gamma$ intensity variation, their extension 
  of 12$^{\prime\prime}$ corresponds to 8.7\,Mm. \label{fig:general}}
\end{figure}

We improve previous sequential measurements by simultaneous observations of seven 
emission lines in a prominence using the THEMIS telescope on Tenerife and its echelle spectrograph. 
The prominence at the west limb, $52^\circ$N, on June 6, 2022 showed negligible
evolutionary variation (according to images from the GONG data archive) and 
can thus be considered as quiescent. Since slit-jaw images were not available, 
we reconstruct the slit positions in a prominence image from the GONG survey 
using the observed distribution of the H$\gamma$ intensity along each slit 
position (Fig.\,\ref{fig:general}).

Given the different wavelengths of the observed lines, the spectrograph
slit is reasonably aligned along the direction of refraction which varies throughout 
the day. Therefore, a spectrograph slit constantly oriented along the direction of 
refraction will rotate over the prominence. Near solstice, however, its direction remains almost constant 
for a few hours  
\citep[see][]{WiehrEtal2019}.
We observe from 9:06 to 9:24 UT and accordingly keep the slit at an angle of 
$72^\circ$ clockwise from north, 
its width corresponding to 1\farcs 5 ($\approx\,1000$\,km on the Sun). Five cameras 
recorded the emission lines H${\gamma}$, Na\,D$_1$, \ion{Sr}{ii}\,4216\,\AA{}, 
and the neighboring lines \ion{He}{i}\,5015\,\AA{} and \ion{Fe}{ii}\,5018\,\AA{}, and 
\ion{Fe}{ii}\,5169\,\AA{} and Mg\,b$_2$\,5172\,\AA{}. These lines are chosen to cover 
an appropriate range of optically thin metal emissions from ions and neutrals  
that matches the grating orders. 

We observe a He-singlet line to avoid correction for overlapping triplet components 
and prefer the slightly weaker Na\,D$_1$ line since D$_2$ has a terrestrial H$_2$O blend 
which does not disappear with the aureole subtraction 
\citep[see][]{WiehrEtal2019}.
\ion{Sr}{ii}\,4216\,\AA{} replaces the formerly used 
\ion{Sr}{ii}\,4078\,\AA{} to fit the grating orders, besides, it avoids 
possible blending with  \ion{Cr}{ii} and \ion{Ce}{ii} lines. 
H${\gamma}$ is a suitable Balmer line of moderate optical thickness 
\citep{Gouttebroze_etal1993}
even in bright prominences required to detect weak metall lines. 
Characteristics of the lines are given in Table\,\ref{tab:Tab1}.

\begin{table*}
  \centering
  \begin{tabular}{|l|c|c|c|c|c|c|c|c|}
    \hline 
    Emitter & $\mu^{-1}$ & $\lambda$ & $\chi _\mathrm{ion}^1$ & $\chi _\mathrm{ion}^2$ &  
    t$_{em}$ & mean E$^\mathrm{tot}$ & mean $\Delta\lambda\,\lambda^{-1}$ & V$_\mathrm{nth}$ \\
    - & - & [\AA] & [eV] & [eV] & [ns] & [erg\,(cm$^2$\,s\,ster)$^{-1}$] & $[10^{-5}]$ & [km\,s$^{-1}$] \\ 
    \hline  
 \ion{Sr}{ii}        & 0.0114 & 4215.52 &  5.7 & 11.0 &   7.1 &   600 & 1.9 & 5.4 \\
 \ion{Fe}{ii}        & 0.0179 & 5018.44 &  7.9 & 16.2 & 500.0 &   200 & 1.4 & 3.8 \\
 \ion{Fe}{ii}        & 0.0179 & 5169.03 &  7.9 & 16.2 & 238.1 &   350 & 1.6 & 4.4 \\
 \ion{Mg}{i}\,b$_2$  & 0.0411 & 5172.68 &  7.6 & 15.0 &  29.7 &   720 & 1.7 & 4.4 \\
 \ion{Na}{i}\,D$_1$  & 0.0435 & 5895.92 &  5.1 & 47.3 &  16.3 &   550 & 1.6 & 4.2 \\ 
 \ion{He}{i}\,singl. & 0.25   & 5015.68 & 24.6 & 54.4 &  74.6 &   330 & 2.4 & 3.8 \\ 
 \ion{H}{i}\,$\gamma$& 1.0    & 4340.72 & 13.6 &  -   & 395.3 &40,000 & 4.6 & 4.5 \\
    \hline
  \end{tabular}
\caption{Line characteristics and data (means of the 4 slit positions). The mean emission
    times t$_{em}$ are obtained as reciprocal values of the transition probabilities taken
    from \citet{Kramida_etal2018}.}
  \label{tab:Tab1}
\end{table*}

We choose four slit positions in the prominence (see Fig.\,\ref{fig:general}) with
a spacing of 2$^{\prime\prime}$. The exposure time at each position is 2\,s.
We repeat the scanning over the four slit positions 50 times. 
The total cadence spans 20\,s (exposure plus moving to the new position). 

In a time interval 
of optimal seeing we average five time steps (no.~$21\,\pm2$) corresponding 
to an effective exposure of 10\,s. Since the line widths do not vary 
significantly over the 50 scans, we apply a running mean over nine camera 
rows corresponding to an effective spatial resolution of 2\farcs 16 
($\approx1500$\,km in the prominence; spatial scale 0\farcs 24 /pixel). 
Disk center spectra are taken for a calibration of the line intensities, 
using the continuum values by Neckel and Labs (1984). The telescope was 
moved around disk center to average solar structures and allow a 
determination of the flat field matrix.

Spectra from the immediate prominence vicinity (long lines in Fig.\,\ref{fig:general}) are 
taken to determine the parasitic light superposed on the emission lines. 
In contrast to strong chromospheric emissions, the metall lines observed here 
are so weak that their emissions are almost lost in the 
aureole light and thus invisible in the raw prominence spectra. Therefore, a 
particularly careful correction is required to ensure that no 
residues of the aureole spectrum remain in the final corrected prominence 
spectra. This procedure is described in detail by 
\citet{Ramelli_etal2012}.

\begin{figure}[ht!]
\includegraphics[width=0.485\textwidth]{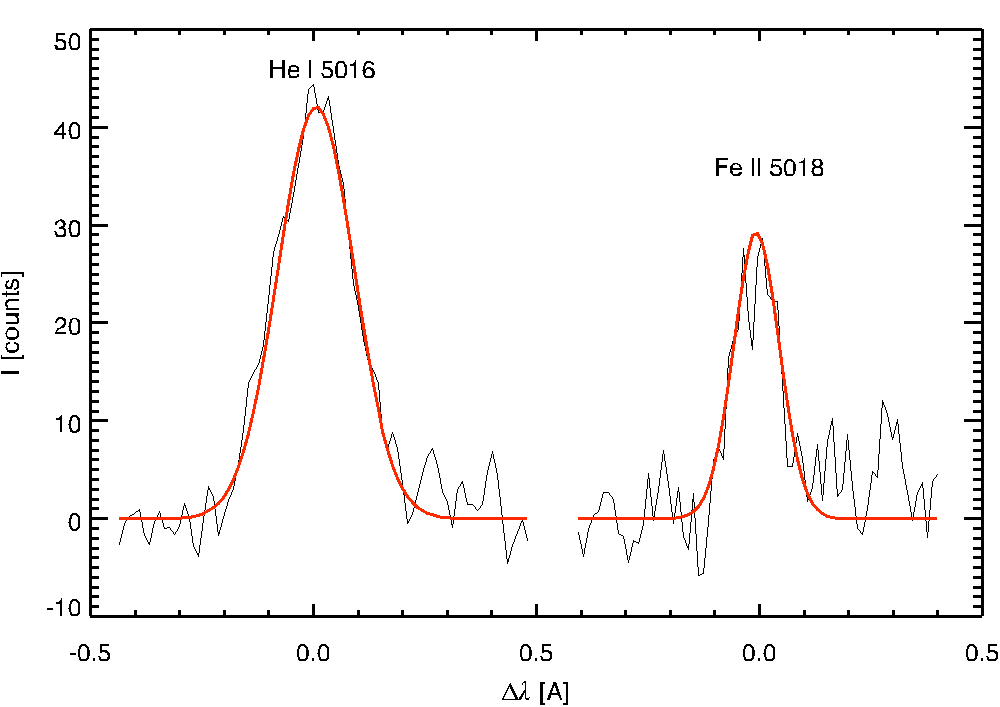}
\caption{Observed line profiles of He-singlet\,5016\,\AA{} and 
\ion{Fe}{ii}\,5018\,\AA{} together with the Gaussian fits. 
Between both lines, $\approx2$\,\AA{} are cut out to stretch the $\lambda$\,scale. 
\label{fig:profiles}}
\end{figure}

The weak metall lines are well represented by fitted Gaussian profiles (Fig.\,\ref{fig:profiles}). 
They are corrected for the spectrograph profile to finally obtain the width of 
each emission line with an accuracy of $\pm3$\,m\AA{}. This is approximately  
$\pm3.5$\% of the width of the metall lines, $\pm2.5$\% of He and $\pm1.5$\% 
of H${\gamma}$. 
Integration of the Gaussian profiles over lambda yields the 
total line emission, $E^\mathrm{tot}$, 
which we consider significant when that of the H{$\gamma$} line amounts 
$E^\mathrm{tot}_{\gamma}>0.4\,E^\mathrm{tot}_\mathrm{max}$ and that of the other emission lines 
$E^\mathrm{tot}>0.75\,E^\mathrm{tot}_\mathrm{max}$, where $E^\mathrm{tot}_\mathrm{max}$ 
is the brightness maximum along the slit for each emission line.

\section{Results} \label{sec:results}

The mean value of the total emission of H${\gamma}$ observed at the four slit positions 
amounts to $E^\mathrm{tot}_{\gamma}= 4\,10^4$ erg\,(s cm$^2$ ster)$^{-1}$, for which 
\citet{Gouttebroze_etal1993}
give $\tau_{\gamma}=0.5$, $\tau_{\alpha}=10.1$ and $E^\mathrm{tot}_{\alpha}=
39.6\,10^4$ erg\,(s cm$^2$ ster)$^{-1}$ (table for $T_\mathrm{kin} = $ 8000\,K,
$P =$ 0.2\,dyn\,cm$^{-2}$ and $\Delta z =$ 5000\,km). 
Such a high Balmer brightness is a necessary 
condition to observe weak optically thin metall lines with sufficient accuracy. 
Since the total emissions of the metall lines are 60\,-\,200 times weaker than 
that of H${\gamma}$ 
they can well be assumed to be optically thin. Indeed, 
\citet{Landman1981}
obtains for $E^\mathrm{tot}_\mathrm{max}$(Na\,D$_1$) = 4420\,erg\,(s cm$^2$ ster)$^{-1}$ 
an optical thickness of $\tau_\mathrm{D}=0.1$. Our Na\,D$_1$ 
emission is 8 times weaker and thus $\tau_\mathrm{D}<0.1$.

\begin{figure}[ht!]
\includegraphics[width=0.475\textwidth]{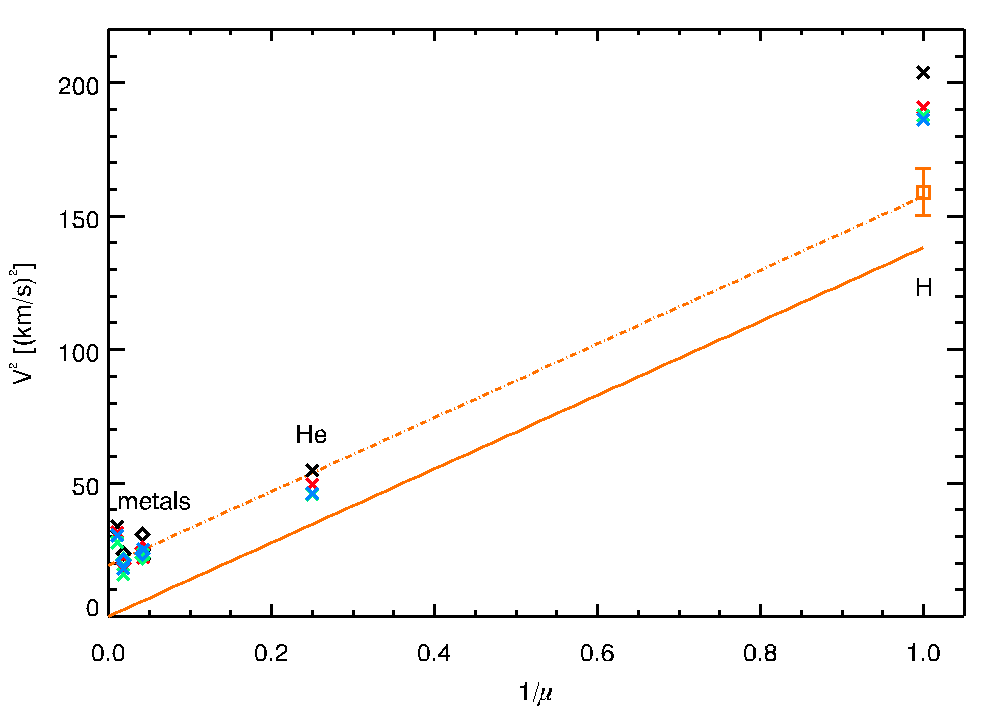}
\includegraphics[width=0.475\textwidth]{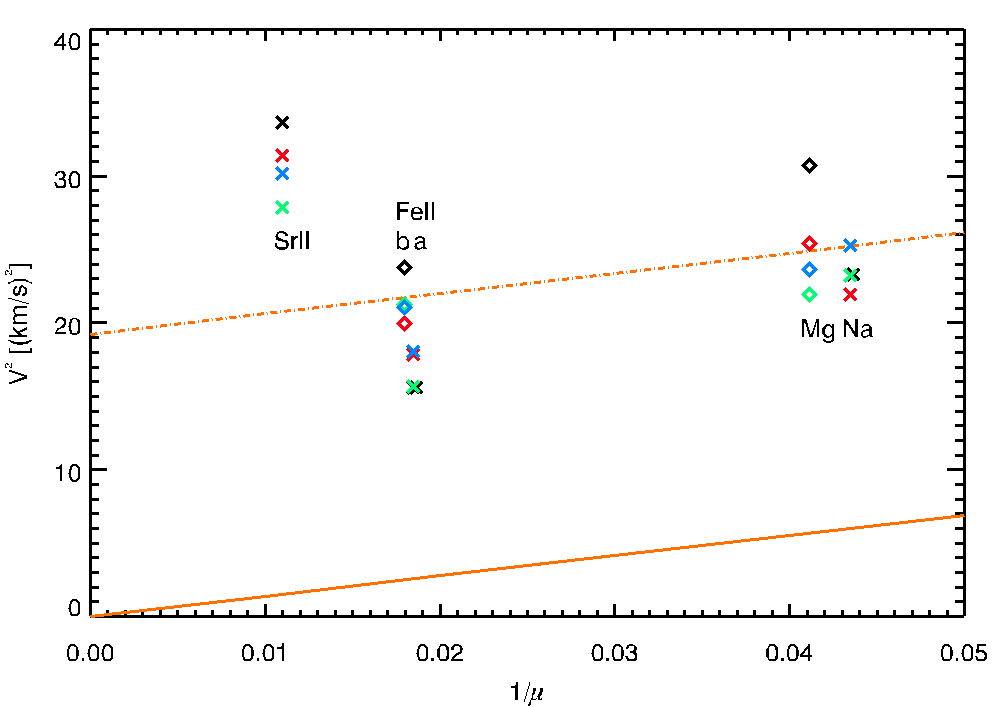}
\caption{Upper panel: observed mean values $V^2=(c\cdot\Delta\lambda_\mathrm{w}/\lambda_0)^2$ 
         versus the inverse atomic mass $1/\mu$ for the 4 slit positions marked in 
         Fig.\,1. The dotted orange line represents a linear fit through all lines, 
         and the solid orange line is used as reference for width excesses of the spectral lines.
         Diamonds for \ion{Fe}{ii}(b) and Mg\,b separate the close 
         \ion{Fe}{ii}(a) and Na\,D values. 
         The orange bar indicates the uncertainty of the opacity broadening of H$\gamma$.
         Lower panel: enlargement of the range of the metall lines. 
         The position of \ion{Fe}{ii}(a) is 
         slightly shifted to separate it from \ion{Fe}{ii}(b).
  \label{fig:my}
  }
\end{figure}

The broadening of optically thin lines is usually described by a thermal and a 
non-thermal term, $V_\mathrm{obs}^2 = V_\mathrm{th}^2 + V_\mathrm{nth}^2$, where $V$ denotes the line 
widths $\Delta\lambda_\mathrm{w}$ in velocity units: $c\cdot\Delta\lambda_\mathrm{w}\,\lambda_0^{-1}$
(c velocity of light, $\lambda_0$ line wavelength,  
$\Delta\lambda_\mathrm{w} = \sqrt(2)\sigma$ the half width at $I_0$e$^{-1}$ 
with $\sigma$ being the width parameter of the Gaussian fits, 
see \citeauthor{Tandberg-Hanssen1995}\,\citeyear{Tandberg-Hanssen1995}). 
We determine the non-thermal
term as difference $V_\mathrm{nth}^2 = V_\mathrm{obs}^2 - V_\mathrm{th}^2$.

In Fig.\,\ref{fig:my}{ }we plot the mean of the observed quantity 
$V^2=(c\cdot\Delta\lambda_\mathrm{w}\,\lambda_0^{-1})^2$ for the four slit positions versus 
the inverse atomic mass $\mu^{-1}$. 
To obtain a reference, we calculate a linear fit through all lines using the mean values over 
the scan positions. The corresponding line is displayed in Fig.\,\ref{fig:my} in dashed orange.
From this fit, we subtract the value at $\mu^{-1} = 0$ and obtain the solid orange line which we use as 
reference for the non-thermal broadening.
This line represents the pure thermal dependence for 8300\,K.
With respect to this reference, the line widths of metals significantly deviate. 
We find 
for \ion{Sr}{ii}, \ion{Fe}{ii}, and Mg\,b$_2$ rather similar non-thermal 
widths as 
\citet{Ramelli_etal2012}. However, we obtain significant $V_\mathrm{nth}$ for Na\,D$_1$, 
which those authors did not find for Na\,D$_2$. We attribute this discrepancy to 
the terrestrial H$_2$O blend in the Na\,D$_2$ wing which diminishes the line width 
\citep[see][]{WiehrEtal2019}. 
H$\gamma$ and \ion{He}{i} also appear above the reference line. For H$\gamma$ one has to take into account 
an additional broadening due to opacity, which we estimate from the tables by \citet{Gouttebroze_etal1993}
for 8000\,K to be (10\,$\pm$\,3)\%. The reduced $V$ is given by the orange bar in Fig.\,\ref{fig:my}.

\begin{figure}[ht!]
\includegraphics[width=0.48\textwidth]{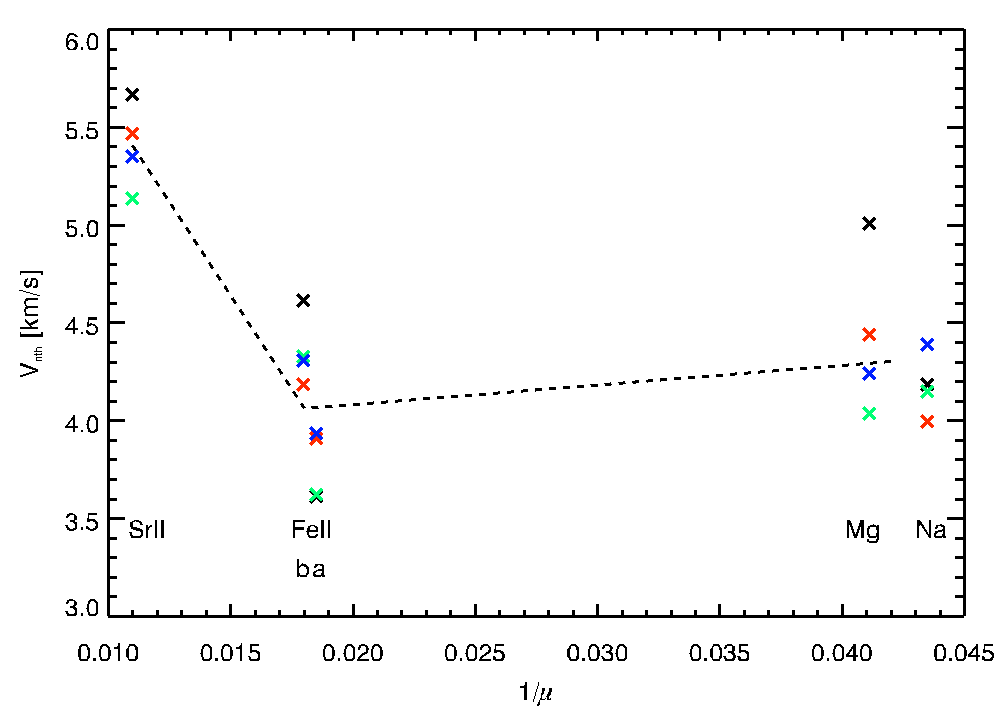}
\caption{Deviation of the observed widths of metall lines from the reference line in Fig.\,\ref{fig:my}
    in velocity units $c\cdot\Delta\lambda_\mathrm{w} 
    \lambda_0^{-1}$ [km/s]. Values for the 4 slit positions appear 
    in the same colors as in Fig.\,\ref{fig:general}. \ion{Fe}{ii}(a)\,5018\,\AA{} is shifted by 
    $\Delta 1/\mu =0.005$ to separate from \ion{Fe}{ii}(b)\,5169\,\AA. }
    \label{fig:excess}
\end{figure}

In Fig.\,\ref{fig:excess}{ } 
we plot for the metall lines the deviations of the observed widths from the reference line in 
Fig.\,\ref{fig:my} 
in velocity units 
$c\,\Delta\lambda_\mathrm{w}\,\lambda_0^{-1}$ versus the inverse atomic mass $\mu^{-1}$ 
(as in Fig.\,\ref{fig:my}) and find a range of $3.5\,<\,V_\mathrm{nth}\,<\,6.0$\,km\,s$^{-1}$.
Mean values are listed in the last column of Tab.\,\ref{tab:Tab1}.

\begin{table*}
  \centering
  \begin{tabular}{|l|c|c|c|c|c|c|}
   \hline 
   Emitter & $l_\mathrm{fp}$ & $t_\mathrm{c}$(v=2\,km\,s$^{-1}$) & $t_\mathrm{c}$(v=4\,km\,s$^{-1}$) &   $t_\mathrm{c}$(v=6\,km\,s$^{-1}$) &  $t_\mathrm{c}$(v=8\,km\,s$^{-1}$) & $t_\mathrm{c}$(v=10\,km\,s$^{-1}$)  \\
    \hline  
 \ion{Sr}{ii}        & 11.4\,m &  5.7\,ms &  2.8\,ms &  1.9\,ms &  1.4\,ms &  1.1\,ms \\
 \ion{Fe}{ii}        & 25.2\,m & 12.6\,ms &  6.3\,ms &  4.2\,ms &  3.2\,ms &  2.5\,ms \\
 \ion{Mg}{i}\,b$_2$  & 10.9\,m &  5.4\,ms &  2.7\,ms &  1.8\,ms &  1.3\,ms &  1.1\,ms \\
 \ion{Na}{i}\,D$_1$  &  7.2\,m &  3.6\,ms &  1.8\,ms &  1.2\,ms &  0.9\,ms &  0.7\,ms \\ 
 \ion{He}{i}\,singl. & 60.4\,m & 30.2\,ms & 15.1\,ms & 10.0\,ms &  7.6\,ms &  6.0\,ms \\ 
 \ion{H}{i}\,$\gamma$& 37.9\,m & 19.0\,ms &  9.5\,ms &  6.3\,ms &  4.7\,ms &  3.8\,ms \\
    \hline
  \end{tabular}
\caption{Free path lengths and mean collision times for the different ions assuming the 
hydrogen density from the table for $T_\mathrm{kin} = 8000$\,K and $p = 0.2\,\mathrm{dyn\,cm}^{-2}$ 
by \citet{Gouttebroze_etal1993}. }
  \label{tab:Tab2}
\end{table*}

\section{Discussion and Conclusions} \label{sec:discuss}

The non-thermal  line broadening reflects small-scale motions
which, in contrast to  
line shift data, are independent of the line-of-sight angle and supposed to result 
from interaction with the magnetic field 
\citep{Parenti_etal2024}.
The observation of weak metall lines allows to study the bright prominence 
body without influences of optical thickness. The finally obtained line width 
excesses are thus not restricted to the prominence border as for strong 
chromospheric lines. Our lines, however, are too weak for a study 
of the periphery but give sufficient signal only at the bright prominence 
center. 

The most prominent result of our study is the almost equal broadening excess 
of lines from ionized Fe and neutral Na and Mg. For Mg\,b$_2$ this result 
confirms the findings by 
\citet{Ramelli_etal2012}
who argue that triplet 
systems are populated via ionization and recombination. During the 
intermediate state as ion magnetic influences   
modify the velocity distribution. 
We interprete the broadening of neutral metals by a kind of 'ionization memory' occuring 
after recombination 
if the line emission happens before a collision changes the momentum of the atom.
To get an idea about collision times in the prominence plasma, 
we use the equation for the length of the free path length
$$ l_\mathrm{fp} = (4\pi \sqrt(2))^{-1} n^{-1} r^{-2}, $$
where n is the particle density and r the mean radius of the colliding atoms
\citep[][p. 64]{Gerthsen}.
The particle density of hydrogen is taken from 
\citet{Gouttebroze_etal1993}
for parameters $T$=8000\,K, $p$=0.2\,dyn\,cm$^{-3}$, $\Delta z$ = 5000\,km
which fit our observed $E^\mathrm{tot}_\gamma$ (see Tab.\,\ref{tab:Tab1}) 
Atomic radii are given by \citet{Clementi1967}, and 
we take ionic radii from \textit{www.webelements.com}.
These values concern the ground state, but since emitting atoms are excited, we enlarge
radii by a factor of two as coarse estimate. 
We consider collisions of the investigated metals with neutral hydrogen and helium and 
neglect all other species because they are less frequent, and protons are much smaller, 
thus they contribute only little.
Dividing this length by the typical velocity of the colliding particles, we obtain 
the mean collision time $t_\mathrm{c}$. Such velocities vary between 4 and 10\,km\,s$^{-1}$ 
for metals (see Fig.\,\ref{fig:my}), 
resulting in $10^{-3}\,\mathrm{s} < t_\mathrm{c} < 10^{-1}\,\mathrm{s}$.
A detailed overview is given in Table\,\ref{tab:Tab2}.
The widths of metall lines indicate that each emitter has its individual
broadening sensitivity, depending on the time interval between 
recombination and line emission.
The neutral metals apparently have a memory of their previous period 
as ions and maintain their velocity distribution after recombination. This 
becomes plausible comparing the mean emission time after recombination
(10$^{-9}$ -- 10$^{-6}$\,s; see \ref{tab:Tab1}) and the mean collision time (10$^{-4}$ -- 10$^{-1}$\,s).
Although the collision times are approximative, the differences of 
a few orders of magnitude to the emission times justifies our suggestion of an
'ionization memory'.

Concerning the large line broadening excess of Na\,D$_1$ (not observed by 
\citeauthor{Ramelli_etal2012}, \citeyear{Ramelli_etal2012}), 
\citet{Landman1981, Landman1983}
argues that sodium is mostly ionized (see ionization potentials in 
Tab.\,\ref{tab:Tab1}). 
Short time after recombination the D lines are emitted. The mean time span between 
recombination and emission is given in Tab.\,\ref{tab:Tab1}.  
So, magnetic forces had influenced sodium during its previous existence as ion. 

The highest broadening excess is found for \ion{Sr}{ii} in agreement 
with \citet{Ramelli_etal2012}.
Strontium may be particularly sensible to magnetic 
influences since it exists, according to the low second ionization potential in 
Tab.\,\ref{tab:Tab1}, most of the time as \ion{Sr}{iii} 
\citep[see][]{Landman1983}. 
[We note that the \ion{Sr}{ii} line widths are not affected by isotopy 
shifts which, according to \citet{Heilig1961},
amount to about 1\,m\AA.]

Ions experience in the magnetic field the Lorentz force, according to the equation
$$ F = k q v \times B, $$ where $q$ is the electric charge, $B$ the magnetic field, $v$ the velocity, 
and $k$ is a prefactor.
The Lorentz force causes a gyration of an ion around the magnetic fieldlines with the speed of the 
component of $v$
perpendicular to the magnetic field, but not a linear acceleration of the ion. In the 
corresponding plane, all directions occur, and in the consequence we observe a broadening of the 
spectral lines \citep[see also][]{Ballester_etal2018}.
Another broadening effect arises when the velocity component along the magnetic field is non-zero 
and the magnetic field is not everywhere perpendicular to gravity direction, e.~g. in the 
magnetic field configuration suggested by \citet{KippenhahnSchlueter}. Then a pendular motion 
of the ion around the local height minimum of the field lines can occur causing a broadening of 
the emission lines.
Different types of waves also influence the line broadening as well as instabilities or shocks 
and motions in twisted flux tubes.
Such effects are discussed in detail by \citet{Ballester_etal2018}.
Prominences typically consist of thin threads which are not resolved in our data due to our long 
exposure times. These threads can move against each other causing a line broadening.

The observed fact that lines from metall emitters are broader than expected 
from our reference line
(see Fig.\,\ref{fig:my}) suggests that the metall emissions are 
broadened by non-thermal effects.
The observed mean \ion{Sr}{ii} line width would 
require unlikely high $T_\mathrm{kin}=175,000$\,K if only thermally broadened.
Since the observed line profiles are almost perfect Gaussians, we can assume a
Maxwellian distribution of non-thermal motions. 
The helium line also is broader than our reference line indicating a non-thermal broadening. 
This can be due to collisions on a longer time scale than the mean emission time. 
Because of the high ionization potential, only a small fraction of helium atoms will be ionized
\citep[see][]{LabrosseGouttebroze2001}.
The excitation of the helium singlet system is then caused mainly by absorption of EUV radiation 
($\lambda\lambda$ 584\,\AA\, and 537\,\AA) 
from the chromosphere, in the first case then emission of a 2\,$\mu$m photon and absorption of a
photospheric photon ($\lambda\lambda$\,5016\,\AA). In contrast, for the triplet system, 
photo-ionization and recombination play an important role.

Detailed physical processes behind such non-thermal broadening remain still 
unclear. They are assumed to be of magnetic origin \citep{Parenti_etal2024}.
Instabilities at the prominence-corona border can be excluded because our data 
refer exclusively to the prominence bulge. Since lines from neutral emitters seem 
to be broadened according to their intermediate existence as ion, a description 
by a dual scenario 'charged-uncharged' will not be sufficient.

\begin{acknowledgements}
We are indepted to M. Verma for carefully reading and commenting the manuscript.
We thank B. Gelly and D. Laforgue for assistance with the instrument. 
THEMIS is operated by CNRS-INSU in the Spanish Observatorio del Teide, 
Tenerife. E.W. and H.B. received financial support from the European Union’s 
Horizon 2020 research and innovation program under grant agreement No.
824135 (SOLARNET).
\end{acknowledgements}

\bibliographystyle{aa}
\bibliography{prot2022}

\end{document}